\documentclass{article}
\usepackage{spconf,amsmath,graphicx}

\usepackage{cite}
\usepackage{bm}
\usepackage{enumerate}
\usepackage{balance}
\usepackage{float}
\usepackage{array}
\usepackage{multirow}
\usepackage{graphicx}
\usepackage{url}
\usepackage{color}
\usepackage{amssymb}
\usepackage{amsfonts}
\usepackage{amsmath}
\usepackage{extarrows}
\usepackage{graphicx}
\usepackage{amsfonts}
\usepackage{caption}
\usepackage{algorithm}
\usepackage{algpseudocode}
\usepackage{indentfirst}
\usepackage{caption}
\usepackage{esint} 
\usepackage{graphicx}
\usepackage{graphics}
\usepackage{pgfplots}
\usepackage{tikz}
\usepackage{epsfig}
\usepackage{soul}
\usepackage{caption}
\usepackage{gensymb} 
\DeclareMathOperator*{\argmin}{argmin}

\def\ba{{\mathbf a}}

\def\bh{{\mathbf h}}

\def\bp{{\mathbf p}}

\def\bw{{\mathbf w}}

\def\by{{\mathbf y}}

\def\bA{{\mathbf A}}
\def\bB{{\mathbf B}}

\def\bI{{\mathbf I}}

\def\bW{{\mathbf W}}

\title{Compressed-Sensing-Based 3D Localization\\ with Distributed Passive Reconfigurable Intelligent Surfaces}
\name{Jiguang He$^{\star}$ \qquad Aymen~Fakhreddine$^{\star}$ \qquad Henk Wymeersch$^{\dagger}$  \qquad George C. Alexandropoulos$^{\star \ddagger}$}
 
  \address{$^{\star}$ Technology Innovation Institute, 9639 Masdar City, Abu Dhabi, United Arab Emirates \\
  $^{\dagger}$ Department of Electrical Engineering, Chalmers University of Technology, Gothenburg, Sweden \\
      $^{\ddagger}$ Department of Informatics and Telecommunications,\\ National and Kapodistrian University of Athens, 15784 Athens, Greece}

\begin{document}
\maketitle
\begin{abstract}
In this paper, the programmable signal propagation paradigm, enabled by Reconfigurable Intelligent Surfaces (RISs), is exploited for high accuracy $3$-Dimensional (3D) user localization with a single multi-antenna base station. Capitalizing on the tunable reflection capability of passive RISs, we present a two-stage user localization method leveraging the multi-reflection wireless environment. In the first stage, we deploy an off-grid compressive sensing approach, which is based on the atomic norm minimization, for estimating the angles of arrival associated with each RIS, which is followed, in the second stage, by a maximum likelihood location estimation initialized with a least-squares line intersection technique. The presented numerical results showcase the high accuracy of the proposed 3D localization method, verifying our theoretical Cram\'er Rao lower bound analysis.   
\end{abstract}

\vspace{-4mm}
\section{Introduction}
\vspace{-2mm}
Reconfigurable Intelligent Surfaces (RISs) \cite{huang2019reconfigurable,huang2019holographic,RISE6G_COMMAG} are expected to play an essential role, not only in wireless communications, but also in radio localization~\cite{wymeersch2019radio,He2019large,Alexandropoulos2022}. For the latter objective, an RIS, irrespective of its sensing capability \cite{Tsinghua_RIS_Tutorial, HRIS_Mag, hardware2020icassp}, can act as an additional node of reference, offering a virtual Line-of-Sight (LoS) path between the Base Station (BS) and User Equipment (UE)~\cite{He2019large}. This feature has motivated the design of RIS-assisted localization systems \cite{Keykhosravi2022infeasible}, which are being optimized to outperform their RIS-free counterparts~\cite{Elzanaty2021}.

In principle, if there exists only a single RIS in a system, it is impossible to perform $3$-Dimensional (3D) localization by only relying on the angular parameters, e.g., Angles of Arrival (AoAs) and/or Angles of Departure (AoDs). In fact, temporal parameters need to be leveraged for localization, whose high precision estimation necessitates wideband operation. In~\cite{Alexandropoulos2022}, multi-RIS-assisted 3D localization was presented for narrowband systems, where each RIS was equipped with a single receive radio-frequency chain enabling reception of pilot signals. The proposed system requires, in principle, backhaul links to connect all RISs to a fusion center, which leads in increased orchestration complexity and deployment cost.   

In this paper, we present a 3D localization narrowband system consisting of a single multi-antenna BS and multiple spatially distributed passive RISs. Differently from~\cite{Alexandropoulos2022}, none of the RISs possesses any radio-frequency chain nor baseband processing capability, thus, being passive with almost zero power consumption. The BS receives the pilot signals from the UE via the RISs, extracts their AoA information, and calculates the UE's 3D position. To separate the received signal reflections from the multiple RISs, the BS applies Zero Forcing (ZF) filtering. Our 3D localization scheme is based on an Atomic Norm Minimization (ANM) framework, which is numerically shown to attain the theoretical Cram\'er Rao Lower Bound (CRLB).

\textit{Notations}: A bold lowercase letter $\ba$ denotes a vector, and a bold capital letter $\bA$ represents a matrix. $(\cdot)^\mathsf{T}$, $(\cdot)^\mathsf{H}$, $(\cdot)^{\dagger}$, and $(\cdot)^{-1}$ denote the matrix transpose, Hermitian transpose, pseudo-inverse, and inverse, respectively. $\mathrm{diag}(\ba)$ is a square diagonal matrix with the entries of $\ba$ on its diagonal, $\mathrm{vec}(\bA)$ denotes the vectorization of $\bA$ by stacking its columns one top of another, $\circ$ and $\otimes$ are the element-wise and Kronecker product operators, respectively, $\mathbb{E}\{\cdot\}$ is the expectation operator, $\mathbf{0}$ denotes the all-zero matrix, $\bI_{M}$ is the $M\times M$ $(M\geq2)$ identity matrix, and $j = \sqrt{-1}$ is the imaginary unit. $\mathrm{Tr}()$ and $\mathrm{Toep}()$ denote the trace operator and a Toeplitz matrix formulated by the argument within the brackets, respectively, $\|\cdot\|_2$ returns the Euclidean norm of a vector and $|\cdot|$ provides the absolute value of a complex number. $\mathcal{CN}(a,b)$ denotes the complex Gaussian distribution with mean $a$ and variance~$b$.
\vspace{-4mm}
\section{System Model}\label{sec:System_Model}
\begin{figure}[t]
	\centering
\includegraphics[width=1 \linewidth]{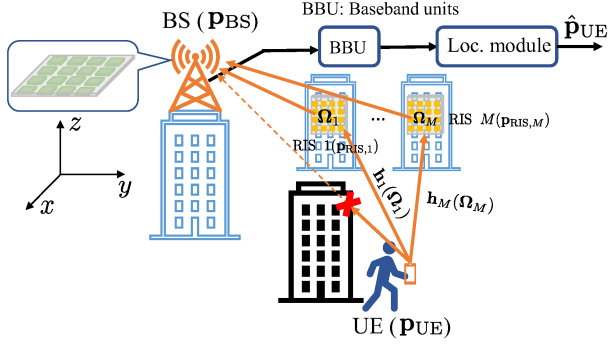}
	\caption{The considered 3D localization system, where the BS localizes the UE in the uplink via the assistance of multiple passive RISs, enabling multiple tunable signal reflections.}
		\label{System_Model}
		\vspace{-0.5cm}
\end{figure}

\vspace{-2mm}
The proposed 3D localization system with distributed passive RISs, which is shown in Fig.~\ref{System_Model}, consists of one $N$-antenna BS, $M$ RISs each with $L$ elements, and a single-antenna UE. Both the BS and RISs include their radiating elements in uniform planar array (UPA) structures. Without loss of generality, the BS is located parallel to the $x\text{-}y$ plane, while the RISs are located parallel to either the $x\text{-} z$ or $y\text{-}z$ planes (e.g., on the facades of buildings or a room's walls). 

\vspace{-4mm}
\subsection{Uplink Sounding}
\vspace{-2mm}
At time slot $t$, the UE transmits the sounding reference signal $s$ with constant transmit power $P$, such that $\mathbb{E}\{|s|^2\} = P$, which reaches the BS via reflections from all RISs. The phase reflection configuration of each $m$-th RIS, with $m=1,2,\ldots,M$, at each $t$-th slot is denoted by $\boldsymbol{\Omega}_{m,t}$. The received signal $\by_{t}\in\mathbb{C}^{N \times 1}$ at the BS for $t=1,2,\ldots,T$, with $T$ being the training overhead, can be expressed as:
\begin{equation}
\by_{t} = \sum^{M}_{m = 1} \underbrace{l_{m}\boldsymbol{\alpha}\left(\boldsymbol{\varphi}_{m}\right)\boldsymbol{\alpha}^{\mathrm{T}}\left(\boldsymbol{\phi}_{m}\right)\boldsymbol{\Omega}_{m,t}\boldsymbol{\alpha}\left(\boldsymbol{\theta}_m\right)}_{\bh_m(\boldsymbol{\Omega}_{m,t})}s+\mathbf{n}_{t},
\label{eq:ObservedSignalBS}
\end{equation}
where $l_{m}\triangleq\sqrt{g^\text{U-R}_{ m}g^\text{R-B}_{ m}} \exp{(j\nu_{m})}$ includes the gain of the end-to-end (e2e) signal propagation path with $g^\text{U-R}_{ m}\triangleq\lambda^2/\left(4\pi d^\text{U-R}_{m}\right)^2$ being the free-space pathloss between the UE and the $m$-th RIS and $g^\text{R-B}_{ m}\triangleq\lambda^2/\left(4\pi d^\text{R-B}_{m}\right)^2$ is the free-space pathloss between the $m$-th RIS and the BS~\cite{Keykhosravi2022}. In the latter definitions, $\lambda$ is the signal wavelength, $d^\text{U-R}_{m}$ and $d^\text{R-B}_{m}$ are the UE-to-$m$-th-RIS and $m$-th-RIS-to-BS distances, respectively, and $\nu_{m} \triangleq 2 \pi f_c \tau_m$ with $f_c\triangleq\frac{c}{\lambda}$ being the carrier frequency and $\tau_m$ is the propagation delay associated with the $m$-th RIS, where $c$ is the speed of light. The BS spatial response vector $\boldsymbol{\alpha}(\boldsymbol{\varphi}_{m})\in\mathbb{C}^{N\times 1}$ is as a function of elevation and azimuth AoAs $\boldsymbol{\varphi}_{m}\triangleq\left(\varphi^{\mathrm{el}}_{m},\varphi^\mathrm{az}_{m}\right)$ associated with the $m$-th RIS. The RIS spatial response vectors are expressed as $\boldsymbol{\alpha}(\boldsymbol{\phi}_m)\in\mathbb{C}^{L\times 1}$ and $\boldsymbol{\alpha}(\boldsymbol{\theta}_m)\in\mathbb{C}^{L\times 1}$, where $\boldsymbol{\phi}_m\triangleq\left(\phi^\mathrm{el}_{m},\phi^\mathrm{az}_{m}\right)$ includes the AoDs from the $m$-th RIS and $\boldsymbol{\theta}_m\triangleq\left(\theta^\mathrm{el}_{m},\theta^\mathrm{az}_{m}\right)$ includes the AoAs to the $m$-th RIS. In particular, depending on the reference plane the RIS/BS UPA is parallel to, $\boldsymbol{\alpha}(\cdot)$ is composed as $\boldsymbol{\alpha}(\cdot) = \boldsymbol{\alpha}_a(\cdot) \otimes  \boldsymbol{\alpha}_b(\cdot)$, where $a,b \in \{x,y,z\}$ and $a \neq b$. For instance, with the assumption of half-wavelength inter-element spacing, $\boldsymbol{\alpha}_a(\boldsymbol{\theta}_m ) \triangleq  [1, e^{j \pi \gamma_a },  \ldots, e^{j \pi (L_{a} -1) \gamma_a} ]^{\mathsf{T}}$, with $L_a$ being the number of RIS elements across the $a$-axis and $\gamma_a  = \cos(\theta^\mathrm{az}_{m} ) \sin(\theta^\mathrm{el}_{m})$ when $a=x$; $\gamma_a  = \sin(\theta^\mathrm{az}_{m} ) \sin(\theta^\mathrm{el}_{m})$ for $a=y$, and $\gamma_a  = \cos(\theta^\mathrm{el}_{m})$ when $a=z$. We define the e2e channel associated with each $m$-th RIS as $\bh_m(\boldsymbol{\Omega}_{m,t}) \triangleq l_{m}\boldsymbol{\alpha}\left(\boldsymbol{\varphi}_{m}\right)\boldsymbol{\alpha}^{\mathrm{T}}\left(\boldsymbol{\phi}_{m}\right)\boldsymbol{\Omega}_{m,t}\boldsymbol{\alpha}\left(\boldsymbol{\theta}_m\right)$, which is a function of the RIS phase profile matrix $\boldsymbol{\Omega}_{m,t}\triangleq\mathrm{diag}(\boldsymbol{\omega}_{m,t}) \in \mathbb{C}^{L \times L}$, where $\boldsymbol{\omega}_{m,t}\in \mathbb{C}^{L \times 1}$ denotes the RIS phase configuration at the $t$-th time slot. We consider passive RISs with elements realized via PIN diodes, hence, holds $ |[\boldsymbol{\Omega}_{m,t}]_{ii}| = 1$ $\forall$$i = 1,2,\ldots,L$~\cite{Wu2019}. 
Finally, the vector $\mathbf{n}_{t}\in\mathbb{C}^{N\times 1}$ in \eqref{eq:ObservedSignalBS} is the additive white Gaussian noise, which is distributed as $\mathcal{CN}(\mathbf{0},\rho\mathbf{I}_N)$ with $\rho$ being its variance.  

\vspace{-4mm}
\subsection{RISs' Reflections Separation}
\vspace{-2mm}
By assuming that the BS knows \textit{a priori} the positions of all RISs, the signal received at the BS from the reflection at the $m$-th RIS can be separated by multiplying \eqref{eq:ObservedSignalBS} with the conjugate transpose of a vector $\mathbf{w}_{m} \in\mathbb{C}^{N\times 1}$, which is orthogonal to every $\boldsymbol{\alpha}(\boldsymbol{\varphi}_{n})$ $\forall n\neq m$ with $n=1,2,\ldots,M$. The matrices $\mathbf{A}\triangleq [\boldsymbol{\alpha}(\boldsymbol{\varphi}_{1}),\boldsymbol{\alpha}(\boldsymbol{\varphi}_{2}),\ldots, \boldsymbol{\alpha}(\boldsymbol{\varphi}_{M})]\in\mathbb{C}^{N\times M}$ and $\mathbf{W}\triangleq [\mathbf{w}_{1},\mathbf{w}_{2},\ldots,\mathbf{w}_{M}]\in\mathbb{C}^{N\times M}$ are hence such that $\mathbf{A}^{\mathrm{H}}\mathbf{W}=\mathbf{I}_M$. Then, $\mathbf{W}$ can be easily computed as the pseudo-inverse of $\mathbf{A}^{\mathrm{H}}$, i.e., $(\bA^{\mathrm{H}})^{\dagger}$. After applying this ZF technique, the post-processed signal received at the BS via the reflection from the $m$-th RIS is given by:
\begin{equation}
z_{m,t} = l_{m}\boldsymbol{\alpha}^{\mathrm{T}}\left(\boldsymbol{\phi}_{m}\right)\mathbf{\Omega}_{m,t}\boldsymbol{\alpha}\left(\boldsymbol{\theta}_m\right)s+\mathbf{w}^{\mathrm{H}}_{m}\mathbf{n}_{t}.  
\label{eq:ZF_Signal}
\end{equation}
Collecting all $T$ consecutive $z_{m,t}$'s into the vector $\mathbf{z}_{m}\triangleq [z_{m,1},z_{m,2}, \ldots,z_{m,T}]^\mathsf{T}\in\mathbb{C}^{T\times 1}$, yields the expression:
\begin{equation}\label{bz_m}
\mathbf{z}_{m} = l_{m}\boldsymbol{\Omega}^{\mathrm{T}}_{m}\left(\boldsymbol{\alpha}\left(\boldsymbol{\phi}_{m}\right)\circ \boldsymbol{\alpha}(\boldsymbol{\theta}_{m})\right) s+\tilde{\mathbf{n}}_{m},
\end{equation}
where $\boldsymbol{\Omega}_{m}\triangleq [\boldsymbol{\omega}_{m,1},\boldsymbol{\omega}_{m,2},\ldots,\boldsymbol{\omega}_{m,T}]\in\mathbb{C}^{L\times T}$ and $\tilde{\mathbf{n}}_m\triangleq [\mathbf{w}^{\mathrm{H}}_{m}\mathbf{n}_{1} ,\mathbf{w}^{\mathrm{H}}_{m}\mathbf{n}_{2},\ldots,\mathbf{w}^{\mathrm{H}}_{m}\mathbf{n}_{T}]^\mathsf{T}\in\mathbb{C}^{T\times 1}$. It can be easily concluded that $\mathbb{E}\{ \tilde{\mathbf{n}}_m\} = \mathbf{0}$ and $\mathbb{E}\{\tilde{\mathbf{n}}_m \tilde{\mathbf{n}}_m^\mathsf{H} \} = \rho \|\bw_m\|_2^2 \bI_T$.

It is noted that the considered ZF combining affects the additive noise term in~\eqref{bz_m}, which now becomes colored. Hence, to guarantee acceptable yet balanced performance for the channel parameter estimation associated with each RIS-enabled signal propagation path, we need to ensure that $\|\bw_1\|_2 \approx \|\bw_2\|_2 \approx \ldots \approx \|\bw_M\|_2$. This condition is generally affected by the number of BS antennas, RIS placement, and BS location. 

\textbf{Remark 1}: In~\eqref{bz_m}, $\boldsymbol{\Omega}_{m}$ acts as the measurement matrix in the compressive sensing nomenclature. Its random configuration can be an intuitive, yet suboptimal, way to guarantee the maximization of its \textit{mutual incoherence}, or equivalently, the satisfaction of the \textit{Restricted Isometry Property (RIP)}.   

\vspace{-4mm}
\section{Proposed Localization Approach}\label{sec:3d_loc}
\vspace{-2mm}
\subsection{ANM-Based AoA Estimation}
\vspace{-2mm}
The problem in~\eqref{bz_m} turns out to be a sparsity-one signal recovery problem. According to the RIS deployment method in Section~\ref{sec:System_Model}, the component of the sparse signal is in the form $\boldsymbol{\alpha}\left(\boldsymbol{\phi}_{m}\right)\circ \boldsymbol{\alpha}(\boldsymbol{\theta}_{m}) = (\boldsymbol{\alpha}_a(\boldsymbol{\phi}_{m} ) \otimes \boldsymbol{\alpha}_z(\boldsymbol{\phi}_{m} ) ) \circ  (\boldsymbol{\alpha}_a(\boldsymbol{\theta}_{m} ) \otimes \boldsymbol{\alpha}_z(\boldsymbol{\theta}_{m} ) ) =(\boldsymbol{\alpha}_a(\boldsymbol{\phi}_{m} ) \circ \boldsymbol{\alpha}_a(\boldsymbol{\theta}_{m} )) \otimes (\boldsymbol{\alpha}_z(\boldsymbol{\phi}_{m} ) \circ \boldsymbol{\alpha}_z(\boldsymbol{\theta}_{m} ))$ for $a \in \{x,y\}$, which is a Kronecker product of two array response vectors. Thus, we define the following atomic set \cite{yang2016,he2020anm}:
\begin{equation}\label{atomic_set}
\mathcal{A} \triangleq \{\boldsymbol{\alpha}_a(x_1, x_2) \otimes \boldsymbol{\alpha}_z(x_3), x_1,x_2,x_3 \in [-\pi, \pi] \}, 
\end{equation}
where $a \in \{x,y\}$. We further introduce the vector $\tilde{\bh}_m \triangleq l_{m}\left(\boldsymbol{\alpha}\left(\boldsymbol{\phi}_{m}\right)\circ \boldsymbol{\alpha}(\boldsymbol{\theta}_{m})\right) = \eta \boldsymbol{\alpha}_a(\tilde{x}_1, \tilde{x}_2) \otimes \boldsymbol{\alpha}_z(\tilde{x}_3) $, where $\eta >0$ and $\exists  \tilde{x}_1,\tilde{x}_2,\tilde{x}_3 \in [-\pi, \pi] $, which is linear combination of the atom(s) belonging in the atomic set $\mathcal{A}$. The atomic norm with respect to $\mathcal{A}$ can be thus written as follows~\cite{yang2016}:
\begin{align}\label{eq:OP}
\|\tilde{\bh}_m\|_{\mathcal{A}} = &\mathrm{inf}_{\mathcal{B}}\Big\{\frac{1}{2L} \mathrm{Tr}(\mathrm{Toep}(\mathcal{U}_m)) + \frac{t_m}{2}\Big\}, \nonumber\\
&\text{s.t.} \;\begin{bmatrix} \mathrm{Toep}(\mathcal{U}_1)  & \tilde{\bh}_m\\
\tilde{\bh}_m^{\mathsf{H}}& t_m
\end{bmatrix} \succeq \mathbf{0},
\end{align}
where set $\mathcal{B}\triangleq\{\mathcal{U}_m \in \mathbb{C}^{L_{a} \times L_{z} }, t_m\in \mathbb{R}\}$ with $\mathcal{U}_m$ being a $2$-way tensor and $\mathrm{Toep}(\mathcal{U}_m)$ is a $2$-level block Toeplitz matrix~\cite{Yang2016IT}. Taking into acount the effect of the noise term in~\eqref{bz_m}, we formulate the following regularized optimization problem (termed as ANM) to find the optimal solution for $\tilde{\bh}_m$:
\begin{align}\label{ANM}
\hat{\tilde{\bh}}_m \triangleq \argmin\limits_{\tilde{\bh}_m \in \mathbb{C}^L,\; \mathcal{B}} &\mu_m \|\tilde{\bh}_m\|_{\mathcal{A}} + \frac{1}{2}\|\mathbf{z}_{m}- \sqrt{P}\boldsymbol{\Omega}^{\mathrm{T}}_{m}\tilde{\bh}_m\|_2^2 \nonumber\\
&\text{s.t.} \;\begin{bmatrix} \mathrm{Toep}(\mathcal{U}_m)  & \tilde{\bh}_m\\
\tilde{\bh}_m^{\mathsf{H}}& t_m
\end{bmatrix} \succeq \mathbf{0},
\end{align}
where $\mu_m \propto \sqrt{\rho} \|\bw_m\|_2 \sqrt{ L \log(L)}$ is the regularization term of the atomic norm penalty. We can construct the covariance matrix of $\tilde{\bh}_m $ based on $\hat{\tilde{\bh}}_m$, and then decompose it to obtain an estimate for the covariance matrix of $\boldsymbol{\alpha}_a(\boldsymbol{\phi}_{m} ) \circ \boldsymbol{\alpha}_a(\boldsymbol{\theta}_{m} )$ and $\boldsymbol{\alpha}_z(\boldsymbol{\phi}_{m} ) \circ \boldsymbol{\alpha}_z(\boldsymbol{\theta}_{m})$. By following the fact that we know $\boldsymbol{\phi}_m$'s (due to the known position of the RISs at the BS) and that $\boldsymbol{\alpha}_a(\boldsymbol{\theta}_{m} ) \boldsymbol{\alpha}_a(\boldsymbol{\theta}_{m} )^\mathsf{H} =  (\boldsymbol{\alpha}_a(\boldsymbol{\phi}_{m} ) \circ \boldsymbol{\alpha}_a(\boldsymbol{\theta}_{m} )) (\boldsymbol{\alpha}_a(\boldsymbol{\phi}_{m} ) \circ \boldsymbol{\alpha}_a(\boldsymbol{\theta}_{m} ))^\mathsf{H} \oslash \boldsymbol{\alpha}_a(\boldsymbol{\phi}_{m} ) \boldsymbol{\alpha}_a(\boldsymbol{\phi}_{m} )^\mathsf{H}$ with $\oslash$ denoting the element-wise division of two matrices with the same size, we can eliminate their contributions in the covariance matrices, and hence, compute estimations for the covariance matrices for $\boldsymbol{\alpha}_a(\boldsymbol{\theta}_{m} )$ and similarly for $\boldsymbol{\alpha}_z(\boldsymbol{\theta}_{m})$ $\forall$$m$. Finally, by using the root MUltiple SIgnal Classification (root MUSIC)
, estimations for $\boldsymbol{\theta}_m$'s, denoted as $\hat{\boldsymbol{\theta}}_m$'s, can be derived.

\vspace{-4mm}
\subsection{Mapping AoAs to UE Location}
\vspace{-2mm}
We apply the least-squares principle for mapping the AoA estimates to the 3D position of UE $\bp_{\text{UE}}$, as follows~\cite{Alexandropoulos2022}:
\begin{equation}\label{LS_Loc}
    \hat{\bp}_{\text{UE}} = \left(\sum_{m = 1}^M \bB_m\right)^{-1} \left(\sum_{m = 1}^M \bB_m \bp_{\text{RIS},m} \right),
\end{equation}
where $\bB_m \triangleq \bI_3 - \hat{\boldsymbol{\xi}}_m \hat{\boldsymbol{\xi}}_m^\mathsf{T}$, $\hat{\boldsymbol{\xi}}_m \triangleq [\cos(\hat{\theta}^\mathrm{az}_{m}) \cos(\hat{\theta}^\mathrm{el}_{m})  ,\\ \sin(\hat{\theta}^\mathrm{az}_{m}) \cos(\hat{\theta}^\mathrm{el}_{m}), \sin(\hat{\theta}^\mathrm{el}_{m})   ]^\mathsf{T}$, and $\bp_{\text{RIS},m}$ denotes the known position of the $m$-th RIS.

\vspace{-4mm}
\section{Position Error Analysis}\label{FIM_CRLB}
\vspace{-2mm}
\subsection{FIM for the AoA Estimation}
\vspace{-2mm}
Let the $4$-tuple vector $\boldsymbol{\eta}_m\triangleq\left[\sqrt{g^\text{U-R}_{ m}}\;\nu_m\;\theta^\mathrm{az}_{m}\;\theta^\mathrm{el}_{m}\right]^\mathrm{T}\in\mathbb{R}^{4\times1}$ include the unknown channel parameters in \eqref{eq:ZF_Signal}. The $4\times4$ Fisher Information Matrix (FIM) for this unknown vector, based on the observations in \eqref{eq:ZF_Signal}, is defined as follows:
\begin{equation}\label{eq:FIM_Initial}
\mathbf{J}\left(\boldsymbol{\eta}_m\right)=\frac{2}{\rho_m}\sum_{t=1}^T\Re\left\{\left(\frac{\partial\mu_{m,t}}{\partial\boldsymbol{\eta}_m}\right)^\mathrm{ H}\frac{\partial\mu_{m,t}}{\partial\boldsymbol{\eta}_m}\right\},
\end{equation}
where $\rho_m \triangleq \rho \|\bw_m\|_2^2$ and the noiseless signal from the $m$-th RIS after ZF filtering can be expressed by the function $\mu_{m,t} \triangleq  \sqrt{g^\text{U-R}_{ m}g^\text{R-B}_{ m}} \exp{(j\nu_{m})} \boldsymbol{\omega}^{\mathrm{T}}_{m,t}\left(\boldsymbol{\alpha}\left(\boldsymbol{\phi}_{m}\right)\circ \boldsymbol{\alpha}(\boldsymbol{\theta}_{m})\right) s$, whose derivative can be computed as follows:
\begin{align}
&\frac{\partial {\mu}_{m,t}}{\partial\boldsymbol{\eta}_m}=\nonumber\\  &\begin{bmatrix}\sqrt{g^\text{R-B}_{ m} }\exp{(j\nu_{m})}\boldsymbol{\omega}^{\mathrm{T}}_{m,t}\left(\boldsymbol{\alpha}\left(\boldsymbol{\phi}_{m}\right)\!\circ\! \boldsymbol{\alpha}(\boldsymbol{\theta}_{m})\right)\! s\\
j\sqrt{g^\text{U-R}_{ m}}\sqrt{g^\text{R-B}_{ m}}\exp{(j\nu_{m})}\boldsymbol{\omega}^{\mathrm{T}}_{m,t}\left(\boldsymbol{\alpha}\left(\boldsymbol{\phi}_{m}\right)\!\circ\! \boldsymbol{\alpha}(\boldsymbol{\theta}_{m})\right)\! s\\
\sqrt{g^\text{U-R}_{ m}}\sqrt{g^\text{R-B}_{ m}}\exp{(j\nu_{m})}\boldsymbol{\omega}^{\mathrm{T}}_{m,t}\left(\boldsymbol{\alpha}\left(\boldsymbol{\phi}_{m}\right)\!\circ\! \frac{\partial{\boldsymbol{\alpha}(\boldsymbol{\theta}_{m})}}{\partial{\theta^\mathrm{az}_{m}}}\right)\! s\\
\sqrt{g^\text{U-R}_{ m}}\sqrt{g^\text{R-B}_{ m}}\exp{(j\nu_{m})}\boldsymbol{\omega}^{\mathrm{T}}_{m,t}\left(\boldsymbol{\alpha}\left(\boldsymbol{\phi}_{m}\right)\!\circ\! \frac{\partial{\boldsymbol{\alpha}(\boldsymbol{\theta}_{m})}}{\partial{\theta^\mathrm{el}_{m}}}\right)\! s
\end{bmatrix}\!\!.
\end{align}

\vspace{-4mm}
\subsection{Position Error Bound}
\vspace{-2mm}
The Fisher information in~\eqref{eq:FIM_Initial} can be used for expressing the $3\times3$ FIM for the unknown UE position $\mathbf{p}_{\text{UE}}$ as \cite{kay1993fundamentals}:
{\begin{equation}\label{eq:fim_position}
\mathbf{J}\left(\mathbf{p}_{\text{UE}}\right)=\sum_{m=1}^M\mathbf{T}_m\mathbf{J}\left(\theta^\mathrm{el}_{m},\theta^\mathrm{az}_{m}\right)\mathbf{T}_m^\mathrm{ T},
\end{equation}
where $\mathbf{T}_m\in\mathbb{R}^{3\times2}$ is the Jacobian matrix $\mathbf{T}_m\triangleq[\frac{\partial\theta^\mathrm{el}_{m}}{\partial\mathbf{p}_{\text{UE}}} \frac{\partial\theta^\mathrm{az}_{m}}{\partial\mathbf{p}_{\text{UE}}}]$}. 

The FIM in~\eqref{eq:fim_position} for the unknown UE position $\mathbf{p}_{\text{UE}}$ can be then used for computing the Position Error Bound (PEB) as
\begin{equation}\label{eq:peb}
\mathrm{ PEB}\triangleq\sqrt{\mathrm{ Tr}\left\{\mathbf{J}^{-1}\left(\mathbf{p}_{\text{UE}}\right)\right\}}\leq\sqrt{\mathbb{E}\left\{\left\|\hat{\mathbf{ p}}_{\text{UE}}-\mathbf{p}_{\text{UE}}\right\|^2\right\}},
\end{equation}
which can serve as a performance lower bound for the Root Mean Square Error (RMSE) of the position estimate $\mathbf{\hat p}_{\text{UE}}$ for $\mathbf{p}_{\text{UE}}$ using the proposed localization scheme in Section~\ref{sec:3d_loc}.

\vspace{-4mm}
\section{Numerical Results}
\vspace{-2mm}
In this section, we evaluate the performance of the proposed 3D localization system. Unless otherwise indicated, we have used the system parameters: $M = 3$, $N =100$, $f_c= 28$ GHz, and $L = 4 \times 4$; the coordinates (centroids) of the three RISs were $\bp_{\text{RIS},1} = (0.5  , 1.5 , 2.9)$, $\bp_{\text{RIS},2} = (-0.5,  0.5,  2.7)$, and $\bp_{\text{RIS},3} = (-0.5,  -0.5,  2.5)$. The UE was located at the origin and the BS was placed at the point $(1, 1, 3)$. For the design of $\boldsymbol{\Omega}_m$'s at the RIS,  we have used columns of the Discrete Fourier Transform (DFT) matrix. 

\vspace{-4mm}
\subsection{Effect of the Training Overhead}
\begin{figure}[t]
	\centering
\includegraphics[width=0.90 \linewidth]{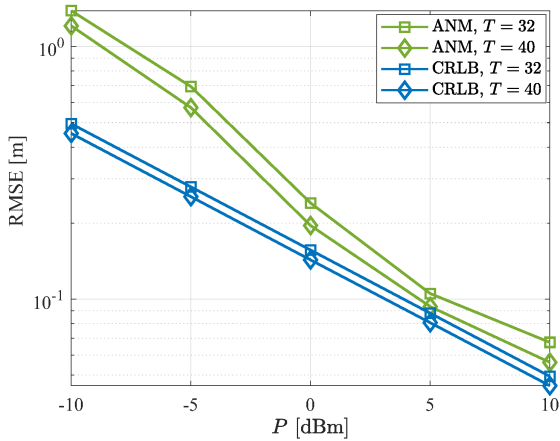}
	\caption{The effect of training overhead $T$ on the performance of the proposed 3D localization system.}
		\label{3D_loc_performance}
		\vspace{-0.5cm}
\end{figure}

\vspace{-2mm}
We herein evaluate the effect of the training overhead $T$ on the proposed 3D localization scheme. The simulation results with different values for $T$ are depicted in Fig.~\ref{3D_loc_performance}, which also includes performance curves for the derived theoretical CRLB in Section~\ref{FIM_CRLB}. It can be seen that, as the transmit power $P$ increases, the positioning accuracy improves, approaching our derived bound. In addition, it is evident that increasing the training overhead from $T = 32$ to $40$ brings only a slight performance enhancement.

Compared to~\cite{Alexandropoulos2022}, we encounter an additional signal propagation hop in our localization system setup. Therefore, more transmit power is required to achieve similar performance. However, by reducing the BS-RIS distance, we can minimize the performance gap between the two systems. Recall that, in our localization system, the spatially distributed RISs are nearly passive, and the BS collects the reflection signals from all RISs simultaneously, without the need of backhaul links and a fusion center, as the system in~\cite{Alexandropoulos2022} requires.

\vspace{-4mm}
\subsection{Effect of the Numbers of BS Antennas and RISs}
\begin{figure}[t]
	\centering
\includegraphics[width=0.90 \linewidth]{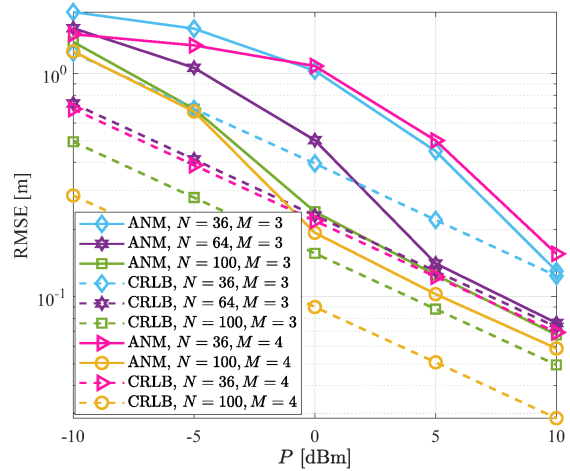}
	\caption{The effect of the number of BS antennas $N$ and that of RISs $M$ on the performance of the proposed 3D localization system.}
		\label{Number_of_BS_antennas_and_RISs}
		\vspace{-0.5cm}
\end{figure}

\vspace{-2mm}
The number of BS antennas $N$ has a significant impact on the design of the ZF combining matrix $\bW$, which in turn affects the proposed 3D localization performance. In Fig.~\ref{Number_of_BS_antennas_and_RISs}, we consider different values for $N$, specifically $N \in \{36, 64,100\}$, while fixing $T = 32$, and illustrate the positioning error of the proposed approach together with the respective theoretical PEB. As shown, increasing $N$ improves both latter metrics and speeds up their performance convergence. We next include another RIS in the system, located at $(1.1, 0.8, 2.8)$, and evaluate its effect on the localization performance. From the CRLB perspective, an obvious gain can be seen by this RIS addition in the setup. However, from the algorithmic perspective, a performance gain is achieved only with a large-sized BS, e.g., for $N = 100$, due to the increased spatial filtering capability at reception. 

\vspace{-4mm}
\subsection{The Role of the RIS Placement}
\begin{figure}[t]
	\centering
\includegraphics[width=0.90 \linewidth]{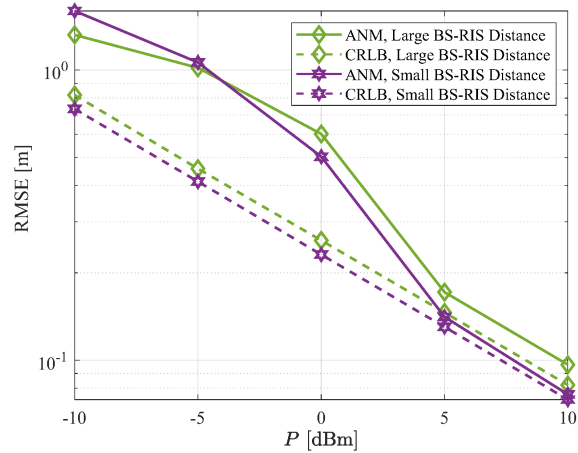}
	\caption{The effect of the BS-RIS distance on the performance of the proposed 3D localization system.}
		\label{Effect_BS_RIS_distance}
		\vspace{-0.5cm}
\end{figure}

\vspace{-2mm}
In this experiment, we move all the RISs $0.86$ m further away from the BS. The simulation results for the localization performance are demonstated in Fig.~\ref{Effect_BS_RIS_distance} for $N = 64$ and $T = 32$. It is obvious that, the larger distance between each RIS and the BS, the worse the performance. The reason lies in that the overall pathloss encountered in each reflection path via the RISs is in this case enlarged. 

\vspace{-4mm}
\section{Conclusion}
\vspace{-2mm}
In this paper, we presented a novel 3D localization system comprising a single multi-antenna BS and multiple spatially distributed passive RISs. Relying on uplink channel sounding and knowledge of the RISs' locations, the BS deploys ZF to simultaneously estimate the AoA feature for all RIS-enabled e2e channels via ANM, which is then used for UE positioning. We showcased the cm-level localization accuracy of the proposed scheme, investigated the impact of various system parameters on its performance, and verified its effectiveness through comparisons with our derived theoretical PEB.

\vspace{-4mm}
\section{Acknowledgement}
\vspace{-2mm}
The work of Profs.~G.~C.~Alexandropoulos and H.~Wymeersch has been supported by the EU H2020 RISE-6G project under grant number 101017011.

\vfill
\pagebreak

\bibliographystyle{IEEEbib}
\bibliography{IEEEabrv,refs}

\end{document}